\documentclass{article}                    % onecolumn
\usepackage{blindtext}
\usepackage{multicol}
\usepackage{gensymb}
\usepackage{amsfonts}
\usepackage{amsmath}
\usepackage{amssymb}
\usepackage{physics}   

\usepackage{threeparttable}
\usepackage{booktabs, caption, makecell}

\usepackage{graphicx}
\usepackage{latexsym}
\usepackage{textgreek}
\usepackage{pifont}
\usepackage{yfonts}
\usepackage{textcomp}
\usepackage[T1]{fontenc}
\usepackage[OT2,OT1]{fontenc}
\usepackage{marvosym}
\usepackage{mathrsfs}

\usepackage[hyphens]{url} % 'hyphens' option allows line breaks after "-" characters
\usepackage{hyperref} % optional

\usepackage{lipsum}
\usepackage{mathtools}
\usepackage{cuted}

\usepackage{color}

\makeatletter
\def\ps@pprintTitle{%
  \let\@oddhead\@empty
  \let\@evenhead\@empty
  \def\@oddfoot{\footnotesize{\it To be submitted to: \@journal}\normalsize \qquad \qquad \qquad \qquad  \quad \,\,\,\,\thepage\hfil\footnotesize{\it\today}}
  \let\@evenfoot\@oddfoot
}
\makeatother

\newcommand{\lshad}{[\![}
\newcommand{\rshad}{]\!]}

\newcommand{\rd}{\textrm{d}}

\newcommand{\vn}{\textrm{\Large{$\upsilon$}}}
\newcommand{\be}{\begin{equation}}
\newcommand{\en}{\end{equation}}

\DeclareMathAlphabet{\mathpzc}{OT1}{pzc}{m}{it}
\begin{document}
\begin{center}
{\Large{\bf{On the structure of isothermal acoustic shocks under classical and artificial viscosity laws: Selected case studies\footnote{DISTRIBUTION A (Approved for public release; distribution unlimited.)}}}}
\end{center}

\begin{center}
{ \large{\bf Sandra Carillo}\\ \normalsize{ Dipartimento Scienze di Base e Applicate per l'Ingegneria, \\
\textsc{Sapienza}  Universit\`a di Roma, ROME, Italy},\\ \& \\{ I.N.F.N. - Sezione Roma1
Gr. IV - MMNLP %Mathematical Methods in NonLinear Physics.
\\ \medskip
 {\large{\bf Pedro  M.\ Jordan}\\ \normalsize{Acoustics Division, U.S.\ Naval Research Laboratory, \\ Stennis Space Center, MS 39529, USA}}}
}\end{center}

\begin{abstract}

Assuming Newton's law of cooling,  the propagation and structure of  isothermal acoustic shocks are studied under 
four different viscosity laws.  Employing both analytical and numerical methods, 1D traveling wave solutions for the velocity and density fields are derived and analyzed.  For each viscosity law considered, expressions for both the shock thickness and the asymmetry metric are determined. And, to ensure that isothermal flow is  achievable,   upper bounds on the associated Mach number values are derived/computed using  the isothermal version of the energy  equation. 
\end{abstract}

\noindent
{\bf Keywords}: {Artificial viscosity, Dispersed shocks,  Isothermal propagation, \\  Newton's law of cooling\, Traveling wave solutions}
% \PACS{PACS code1 \and PACS code2 \and more}
% \subclass{MSC code1 \and MSC code2 \and more}

\section{Introduction}\label{sect:Intro}

In a celebrated paper published in 1849, Stokes~\cite{Stokes49} appears to have been the first to consider the impact of  shear viscosity on the 1D propagation of linear acoustic waves in gases.  Two years later,  Stokes~\cite{Stokes51} went on to examine the propagation of linear, time-harmonic, acoustic  plane waves in a gas whose only loss mechanism is its ability to radiate heat to its surroundings, the process of which he modeled using Newton's law of cooling; see  Appendix~\ref{App:A}. In 1910, Rayleigh~\cite[pp.~270--271]{Rayleigh10}  presented a partial analysis of the isothermal propagation  of an infinite (1D) ``wave of condensation'' in a constant viscosity, but thermally non-conducting, gas, the process of radiation being invoked to maintain the flow's isothermal nature.  Later, Lamb (see, e.g., Ref.~\cite[\S\,360]{Lamb45}) and, in 1953, Truesdell~\cite[p.~687]{Truesdell53} generalized Stokes'~\cite{Stokes51} radiant propagation model  to include the effects of both viscosity and thermal conduction. In 2004, LeVeque~\cite{LeVeque04} employed the assumption of isothermal flow to investigate 1D ``delta shocks'' in lossless perfect gases;  because he did not consider the energy equation, however, LeVeque's analysis, like Rayleigh's~\cite[pp.~270--271]{Rayleigh10}, must be considered as only  partially complete.

The primary aim of the present study is to not only complete, but also  extend Rayleigh's analysis, wherein only the equation of motion (EoM) for the  density field was derived and integrated, by considering the effects of various viscosity laws on the propagation and structure of  acoustic shocks in the setting of the isothermal piston problem.  Specifically, we determine and analyze 1D traveling wave solutions (TWS)s, under the assumption that the temperature of the gas in question is held constant, based on the following four viscosity laws: (i)~constant shear viscosity (i.e., the case considered in Ref.~\cite[pp.~270--271]{Rayleigh10}), (ii)~shear viscosity proportional to  mass density, (iii)~von Neumann--Richtmyer artificial viscosity, and (iv)~Evans--Harlow--Longley artificial viscosity. These particular laws were selected primarily because all  lead to model systems that are amiable to study by analytical means.

To provide a  mechanism  for achieving isothermal flow, we, like Rayleigh, invoke the process of radiation, which like Stokes~\cite{Stokes51} we model via Newton's law of cooling.  To ensure that isothermal flow is not only mathematically but also physically possible, under each of the four cases considered, we also determine, based on the \emph{full} form of the isothermal energy equation, the upper bound on the range of (piston) Mach number values corresponding to each case.

In the next section, we begin our investigation  with a review of the thermodynamics of perfect gases and the formulation of our governing system of equations.\\

\noindent
\textbf{Remark~1}:
As  Chandrasekhar~\cite[Chap.~II]{Chand67} notes,  isothermal flow is a particular type of  polytropic  process---one quantified by   $\gamma^{\prime}=1$, $n= \infty$, where $\gamma^{\prime}$, $n$ are used by Chandrasekhar  to represent the polytropic exponent and polytropic index, respectively; see Ref.~\cite[pp.~41--44]{Chand67}.   

\section{Formulation of mathematical model}\label{sect:2}

\subsection{Thermodynamical aspects  of perfect gases}

As defined by Thompson~\cite{T72}, a \emph{perfect gas} is one in which $p(>0)$, the thermodynamic pressure, $\rho(>0)$, the mass density, and $\vartheta (> 0)$, the absolute temperature,  obey the following special case of the ideal gas law~\cite[\S\,2.5]{T72}:
\be\label{eq:EoS_ideal}
  p =c_{\rm v} (\gamma-1)\rho\vartheta   \qquad (c_{\rm p}, c_{\textrm{v}} := \text{const.}).
\en
Here, $c_{\rm p} > c_{\rm v} > 0$ are the specific heats at constant pressure and constant volume, respectively, and $\gamma =c_{\rm p}/c_{\textrm{v}}$, where $\gamma \in (1,5/3]$ in the case of perfect gases.   Furthermore, a zero (``0'') subscript attached to a field variable identifies the \emph{uniform} equilibrium state value of that variable; i.e., in the present investigation,  the gas is assumed to be \emph{homogeneous} when in its equilibrium state.

With  regard to the perfect gas assumption we note that  the equilibrium state values of the adiabatic and isothermal sound speeds in the gas are given by
\be
c_{0}=\sqrt{c_{\rm p} (\gamma-1)\vartheta_{0}} 
\en
 and
\be\label{eq:b_0}
b_{0}= c_{0}\gamma^{-1/2}=\sqrt{c_{\textrm{v}}(\gamma - 1)\vartheta_{0}},
\en
respectively, where we observe that $b_{0}\in (0, c_{0})$.  That is, $c_{0}$ and $b_{0}$ are, respectively, the speeds of infinitesimal-amplitude acoustic signals under the adiabatic and  isothermal assumptions; see, e.g., Ref.~\cite[\S\,278]{Lamb45}, wherein $b_{0}$ is referred to as the  ``Newtonian velocity of sound''.

In this study we will also make use of the thermodynamic axiom known as the \emph{Gibbs relation}~\cite[p.~58]{T72}, which in the case of a perfect gas can be written as
\begin{equation}\label{eq:Gibbs}
\vartheta \,\rd \eta= \rd E - c_{\rm v} (\gamma-1)\vartheta \rho^{-1}\rd \rho \qquad (\text{perfect gases}),
\end{equation}
where $\eta$ is   the specific entropy and   $E$ is the specific internal energy.  Here, we note for later reference the relations
\be\label{eq:E_H}
E =  c_{\rm v}\vartheta =H/\gamma \qquad (\text{perfect gases}),
\en
where $H$ is the specific enthalpy.

\subsection{Navier--Stokes--Fourier system in 1D}\label{sect:NSF-sys}

For the  1D flow  of a perfect gas along the $x$-axis of a Cartesian coordinate system,   the Navier--Stokes--Fourier (NSF) system~\cite[p.~513]{P89}  can, assuming the absence of all body forces and that  the ratio $\mu_{\rm b}/\mu$ is  constant,  be written in the following form:
\begin{subequations}\label{sys:NS_General}
\begin{align}
\frac{\partial \rho}{\partial t}+u\frac{\partial \rho }{\partial x} + \rho\frac{\partial u}{\partial x}  &=0,\label{eq:cont_dim_NS}\\
\rho \left(\frac{\partial u}{\partial t}+ u\frac{\partial u}{\partial x}\right)   + \rho c_{\rm p}\frac{\partial \vartheta}{\partial x} &= \rho \vartheta \frac{\partial \eta}{\partial x} \notag \\
& \!\!\!\!+\vn\frac{\partial }{\partial x}\!\left( \mu \frac{\partial u}{\partial x}\right)\!,\label{eq:mom_dim_NS}\\
\rho\vartheta \left(\frac{\partial \eta}{\partial t} +u\frac{\partial \eta}{\partial x} \right) - \Phi  &=\rho r - \frac{\partial q}{\partial x},\label{eq:EntropyProd_dim_NS}\\
q &=-K\frac{\partial \vartheta}{\partial x},\label{eq:Fourier}\\
c_{\rm v}\ln(\vartheta/\vartheta_0) -c_{\rm v}(\gamma-1)\ln (\rho/\rho_0) &=\eta-\eta_{0},\label{eq:eta_perfect-gas}
\end{align}
\end{subequations}
where $p=p(x,t)$, $\rho = \rho(x,t)$,  $\vartheta = \vartheta(x,t)$, and $\eta=\eta(x,t)$ under this flow geometry.  In Sys.~\eqref{sys:NS_General}, ${\bf u}=(u(x,t),0,0)$ and ${\bf q}=(q(x,t),0,0)$ are, respectively, the velocity and heat flux vectors;     $\mu (> 0)$ is the shear (or dynamic) viscosity; $\vn =  \tfrac{4}{3} +\mu_{\rm b}/\mu$ is the viscosity number, where $\mu_{\rm b}(\geq 0)$ is the bulk viscosity;  $K(>0)$ is the thermal conductivity; and $r$, which carries  units of W/kg (i.e., m$^{2}$/s$^{3}$), represents the external rate of supply of heat per unit mass.  Moreover, $\Phi$, the dissipation function~\cite{T72,P89},  here takes the form
\be\label{eq:Diss_fun_dim}
\Phi = \mu\vn \left(\partial u/\partial x\right)^{2};
\en
Eq.~\eqref{eq:eta_perfect-gas} follows from integrating Eq.~\eqref{eq:Gibbs} after  substituting in  
$E$ from Eq.~\eqref{eq:E_H}; and we note that the pressure  gradient term that would normally have appeared in  Eq.~\eqref{eq:mom_dim_NS} was eliminated  using the relevant 1D special case of the  thermodynamic relation~\cite[p.~71]{T72}
\begin{equation}\label{eq:Ethalpy_p_temp_grad}
\rho^{-1}\nabla p = \nabla  H - \vartheta \nabla \eta.
\end{equation}

In what follows we  consider the \emph{compressive} version of the classic  (1D) piston problem.  That is,  the piston, whose face we take to be thermally insulated, is located at $x=-\infty$ and moving to the right with constant speed $u_{\rm p}(> 0)$ while the gas at $x=+\infty$ is  in its equilibrium state.   In this regard we also observe that, since the piston's motion is strictly compressive, it follows that  $u \in[0, u_{\rm p}]$ and $\partial u/\partial x \leq 0$ will always hold under the assumed flow geometry.  

Additionally, we will  invoke  the assumption
\be\label{eq:Stokes-hyp}
\mu_{\rm b} = 0,
\en
which of course is \emph{Stokes' hypothesis}~\cite{T72,P89}.  Here, we stress that Eq.~\eqref{eq:Stokes-hyp}, which  in the case of air, e.g., is only an approximation~\cite[Table~1.1]{T72}, will not be a  limitation on our analysis.  This  is because  our numerical results will focus exclusively on the (monatomic) gas Ar --- one for which  Eq.~\eqref{eq:Stokes-hyp} has been shown, by both theory and experiment, to  hold \emph{exactly}; see, e.g., Refs.~\cite{ChapCow70,TM80}. 

\subsection{Isothermal piston problem}\label{sect:IsoPP}

Assuming hereafter that the flow is \emph{isothermal}, i.e., $\vartheta \equiv \vartheta_{0}$, we find that  Eq.~\eqref{eq:EoS_ideal} (i.e., our  EoS) and the constitutive relations given in Eqs.~\eqref{eq:Fourier} and~\eqref{eq:eta_perfect-gas}  reduce, respectively,  to
\be\label{eq:EoS_Iso}
 p = b_{0}^{2} \rho =  p_{0}\rho/\rho_{0},
\en
which  is simply Boyle's law~\cite{Rayleigh10};
\be
q=0,
\en
since $\partial \vartheta_{0}/\partial x =0$; and
\be\label{eq:Entropy_isothermal_dim}
\eta-\eta_{0} =-c_{\textrm{v}}(\gamma - 1)\ln(\rho/\rho_{0}),
\en
which is the isothermal special case of Eq.~\eqref{eq:eta_perfect-gas}.

On carrying out these substitutions, followed by the use of 
Eqs~\eqref{eq:cont_dim_NS}, \eqref{eq:Diss_fun_dim}, and~\eqref{eq:Stokes-hyp}, Sys.~\eqref{sys:NS_General} is  reduced to
\begin{subequations}\label{sys:NS_Iso}
\begin{align}
\frac{\partial \rho}{\partial t}+u\frac{\partial \rho }{\partial x} + \rho\frac{\partial u}{\partial x}  &=0,\label{eq:cont_NS_Iso}\\
\rho \left(\frac{\partial u}{\partial t}+ u\frac{\partial u}{\partial x}\right)   + b_{0}^{2} \frac{\partial \rho}{\partial x} - \frac{4}{3}\frac{\partial }{\partial x}\!\left( \mu \frac{\partial u}{\partial x}\right) &=0,\label{eq:mom_NS_Iso}\\
\vartheta_{0}+\frac{b_{0}^{2}}{c_{\rm v} \varkappa_{0}}\frac{\partial u}{\partial x}  - \frac{(4/3)}{c_{\rm v} \varkappa_{0}} \left(\frac{\mu}{\rho}\right)\!\left(\frac{\partial u}{\partial x}\right)^{2} &=  \vartheta_{\rm e},\label{eq:r_NS_Iso}
\end{align}
\end{subequations}
where  our adoption of Stokes' hypothesis yields $\vn = 4/3$.   

In Eq.~\eqref{eq:r_NS_Iso}, we have taken $r$ to be given by the isothermal special case of Newton's law of cooling, viz.:
\be\label{eq:Newton_lc-isothermal}
r=- c_{\rm v} \varkappa_{0} (\vartheta_{0} - \vartheta_{\rm e}),
\en
where $\varkappa_{0} (> 0)$, the ``velocity of cooling''~\cite[p.~307]{Stokes51}, carries units of 1/s; see also Ref.~\cite[pp.~651, 687]{Truesdell53}. Here, $\vartheta_{\rm e}=\vartheta_{\rm e}(x,t)$, the  temperature of the surrounding environment,  is an additional dependent variable that is \emph{controllable} by the experimenter based on  ``output'' from Eq.~\eqref{eq:r_NS_Iso}.   This, of course, is necessary because  the gas must  be able to  radiate away the resulting heat of its compression, at a rate that cannot be assumed constant,  if our assumption of isothermal flow is to be satisfied + maintained. (Note that the ability of the gas to conduct heat plays no role under the isothermal assumption.)  In this regard, we have adopted the assumptions stated by Rayleigh~\cite[p.~28]{Rayleigh96} regarding the physical and thermal characteristics of the enclosing cylinder.

\subsection{Viscosity laws: Classical and artificial}

In this subsection we give the precise statement of the four viscosity laws mentioned in Section~\ref{sect:Intro}. Note that the first two  stem from classical continuum theory while the latter two are of the artificial type.

\begin{enumerate}
\item[(i)] Constant shear viscosity: 
\be
\mu =\mu_{0};
\en
 recall, this is  the case considered in Ref.~\cite[pp.~270--271]{Rayleigh10}; it is also the exact form of $\mu$  for the current (i.e., isothermal) problem under the kinetic theory of gases.
\item[(ii)] Shear viscosity proportional to mass density:
\be
\mu =\nu_{0} \rho;
\en
see Refs.~\cite[\S\,12.11]{ChapCow70} and~\cite{MV12}.

\item[(iii)]  von Neumann--Richtmyer (vNR) artificial viscosity:
\be\label{eq:vNR_mu}
\mu= -\mathfrak{b}_{\rm iii}^{2} \lambda_{0}^{2} \rho \!\left(\frac{\partial u}{\partial x}\right)\!;
\en
see Refs.~\cite{vNR50} and \cite[\S\,V-D-1]{Roache72}.
\item[(iv)] Evans--Harlow--Longley (EHL) artificial viscosity:
\be\label{eq:EH_mu}
\mu= \tfrac{1}{2}\mathfrak{b}_{\rm iv} \lambda_{0} \rho u;
\en
see Refs.~\cite[\S\,V-D-2]{Roache72}, \cite[p.~16]{EH57}, and~\cite[p.~11]{Longley60}.
\end{enumerate}
Here, $\mu_{0}$ and $\nu_{0}=\mu_{0}/\rho_{0}$ represent,  respectively,   the (constant) equilibrium state values of the shear and kinematic viscosity coefficients;  in the case of hard-sphere molecules we have, according to the kinetic theory of gases~\cite[\S\,2.7]{T72},
\be\label{eq:mu-lambda}
\mu_{0}=\tfrac{5}{32}\pi\rho_{0}\bar{c}_{0}\lambda_{0},
\en
an expression which also follows on setting ``$\delta$'' in Ref.~\cite[Eq.~(9b)]{ML49} equal to 
$5\pi/32 \approx 0.491$; we let $\lambda_{0}(>0)$ denote the equilibrium state value of the  molecular mean-free-path (see, e.g., Refs.~\cite[p.~680]{ML49} and~\cite[pp.~95--97]{T72}); the equilibrium state value of the mean molecular speed is given by~\cite[p.~108]{T72} 
\be\label{eq:c-bar}
\bar{c}_{0}=c_{0}\sqrt{8/(\gamma \pi)} = b_{0}\sqrt{8/\pi};
\en   
and  $\mathfrak{b}_{\rm iii}(>0)$  and $\mathfrak{b}_{\rm iv}(>0)$ are  adjustable dimensionless parameters~\cite[\S\,V-D-1]{Roache72}.

Lastly, since our investigation is to be carried out  primarily by analytical methodologies,  and we seek an approach that would allow one to compare/contrast these four cases  in a consistent manner, we have taken  $\Delta x = \lambda_{0}$,  where $\Delta x$ is the spatial mesh increment  used in the usual statements of both the vNR and EHL laws.  And with regard to $\lambda_{0}$  we  record here, for later reference,  the expression
\be\label{eq:MFP_nu0_b0}
\lambda_{0} =\frac{16\nu_{0}}{b_{0}\sqrt{50\pi}},
\en
which  is easily obtained after eliminating $\bar{c}_{0}$ between Eqs.~\eqref{eq:mu-lambda} and~\eqref{eq:c-bar}, and we note  that $16/\sqrt{50\pi} \approx 1.277$.

\section{Traveling wave reduction}

\subsection{Ansatzs and wave variable}

Invoking the traveling wave assumption, we set
\be
u(x,t) =f(\zeta), \quad \rho(x,t)=g(\zeta), \quad \vartheta_{\rm e}(x,t)=\Theta_{\rm e}(\zeta),
\en
where  $\zeta := x - vt$ is the  wave (i.e., similarity) variable, and where   the parameter $v(>0)$ will  be seen to represent the speed of the resulting shocks.  On substituting the above ansatzs into  Sys.~\eqref{sys:NS_Iso}, the latter is reduced to the following system of ODEs:
\begin{subequations}\label{sys:TW_Iso}
\be \label{eq:TWcont}
\frac{\rd}{\rd\zeta} \left(-vg + fg\right) = 0, 
\en
\be \label{eq:TWmom}
g(-v+f)f^{\prime} +b_{0}^{2}g^{\prime}-\frac{4}{3}\frac{\rd }{\rd\zeta}\left(\mu f^{\prime}\right) = 0, 
\en
\be\label{eq:TW_Theta_ext}
\vartheta_{0}+\left(\frac{b_{0}^{2}}{c_{\rm v} \varkappa_{0}}\right)\!f^{\prime}  - \frac{(4/3)}{c_{\rm v} \varkappa_{0}} \left(\frac{\mu}{g}\right)\!\left(f^{\prime}\right)^{2} =  \Theta_{\rm e},
\en
\end{subequations}
where a prime denotes $\rd/\rd \zeta$.  

This system is to be integrated subject to the  asymptotic conditions 
\begin{multline} 
 f \to u_{\rm p}, \quad f^{\prime} \to  0 \qquad  (\zeta \to -\infty),
g \to \rho_{0}, \quad  f, f^{\prime} \to  0 \qquad  (\zeta \to +\infty), 
\end{multline} 
which of course correspond to a shock moving to the \emph{right}; recall that $v >0$.

\subsection{Shock speed and associated ODE}

Using the fact that Eq.~\eqref{eq:TWcont} integrates to 
\be \label{eq:TWg}
 g  = \frac{\rho_{0} v}{v - f}, 
\en
allows us to, in turn,  integrate Eq.~\eqref{eq:TWmom}:
\be \label{eq:TWfg}
-\rho_{0} vf +b_{0}^{2}g- (4\mu/3) f^{\prime} = b_{0}^{2}\rho_{0}, 
\en
where the resulting constants of integration were found to be $\mathcal{K}_{1}=-\rho_{0}v$ and $\mathcal{K}_{2}=\rho_{0}b_{0}^{2}$, respectively.  Now eliminating $g$ between the former and latter equations yields
\be \label{eq:TWmotion_v}
(4\mu/3)(v - f) f^{\prime}=\rho_{0}[(b_{0}^{2}-v^{2})f+vf^2],
\en
i.e., a Riccati type equation. 
Employing the asymptotic conditions once again leads us to consider 
\be\label{eq:Quad}
v^{2} -b_{0}^{2}=u_{\rm p}v,
\en
a quadratic whose only positive root is 
\be\label{eq:v}
v=\frac{u_{\rm p} +\sqrt{u_{\rm p}^2+4b_{0}^{2}}}{2} = \tfrac{1}{2}u_{\rm p}\left(1+\sqrt{1+4 \textrm{Ma}^{-2}}\,\right)\!,
\en
 where   $\textrm{Ma} = u_{\rm p}/b_{0}$ is the piston Mach number.

With these results in-hand, we can reduce Sys.~\eqref{sys:TW_Iso}  to the two-equation system
\begin{subequations}\label{sys:TW-final}
\be \label{eq:TWmotion-final}
(4\mu/3)(v - f) f^{\prime}+\rho_{0}vu_{\rm p}(1 - f/u_{\rm p})f =0,
\en
\be\label{eq:Theta_ext-final}
\vartheta_{0}-\left(\frac{b_{0}^{2}}{c_{\rm v} \varkappa_{0}}\right)\!|f^{\prime}|  -  \frac{(4\mu/3)}{c_{\rm v} \varkappa_{0}} \left(\frac{v - f}{\rho_{0} v}\right)\!\left|f^{\prime}\right|^{2} =  \Theta_{\rm e}.
\en
\end{subequations}
Eq.~\eqref{eq:TWmotion-final} is the associated ODE of our traveling wave analysis; in the next four sections, we shall integrate it, under each of the aforementioned  cases of $\mu$, subject to the wave-front condition $f(0)=\tfrac{1}{2}u_{\rm p}$.

\subsection{Shock thicknesses, $Q$-metric, and  jump amplitudes}

Employing the notation
\begin{equation}
f(\pm \infty):=\lim_{\zeta \to\pm\infty}f(\zeta), \qquad  g(\pm \infty):=\lim_{\zeta \to\pm\infty}g(\zeta),
\end{equation}
 we define the shock thickness of the velocity profile by
 \begin{equation}\label{eq:ell_gen_f}
\ell_{j} :=\frac{f(-\infty) - f(+\infty)}{\max\big{|} f^{\prime}(\zeta)\,\big{|}}= \frac{u_{\rm p}}{|f^{\prime}(\zeta_{j}^{\bullet})|},
\end{equation}
a definition which Morduchow and Libby~\cite[p.~680]{ML49} attribute to  Prandtl, and
that of the density profile by
\begin{equation}\label{eq:l_gen_g}
l_{j} :=\frac{g(-\infty) - g(+\infty)}{\max \big{|} g^{\prime}(\zeta)\,\big{|}}%\\
= \left(\frac{2}{-1+\sqrt{1+4 \textrm{Ma}^{-2}}}\right)\frac{\rho_{0}}{|g^{\prime}(\zeta_{j}^{*})|}.
\end{equation}
Here,  $\zeta=\zeta_{j}^{\bullet}$ and $\zeta=\zeta_{j}^{*}$  denote the   relevant stationary points of $f^{\prime}$ and $g^{\prime}$, respectively, i.e., $f^{\prime\prime}(\zeta_{j}^{\bullet})=0$ and $g^{\prime\prime}(\zeta_{j}^{*})=0$; the subscript $j$   represents the  number [i.e., (i)--(iv)] of the case under consideration; and with regard to Eq.~\eqref{eq:l_gen_g} we note the following:
\be\label{eq:g-prime}
 g^{\prime}(\zeta)  = \frac{\rho_{0} vf^{\prime}}{(v - f)^{2}}= - \frac{3\rho_{0}^{2} v^{2}(u_{\rm p}-f)f}{4\mu (v - f)^{3}}
\en
and
\begin{multline}
g(-\infty) =\frac{\rho_{0}v}{v-u_{\rm p}} = \rho_{0}\left(\frac{1+\sqrt{1+4 \textrm{Ma}^{-2}}}{-1+\sqrt{1+4 \textrm{Ma}^{-2}}}\right)
 >  g(+\infty) = \rho_{0}.
\end{multline}

We also define the $Q$- (or asymmetry) metric
\be\label{eq:Q}
Q(\textrm{Ma})=\frac{\displaystyle\int_{-\infty}^{0}[1-R(\zeta)]\,\rd \zeta}{\displaystyle\int_{0}^{+\infty} R(\zeta)\,\rd \zeta},
\en
where
\be
R(\zeta) =\frac{g(\zeta)-\rho_{0}}{g(-\infty)-\rho_{0}} \implies \lim_{\zeta \to \mp \infty}R(\zeta)=1,0,
\en
respectively.  As Schmidt~\cite[p.~369]{Schmidt69}\footnote{Eq.~\eqref{eq:Q} differs from Schmidt's expression for $Q$ because in Ref.~\cite{Schmidt69} the shocks propagate to the left.} points out, not only is $Q$  a ``sensitive measure of asymmetry'', but it also complements the shock thickness as a characterizing metric since the latter ``fails to give  sufficiently detailed information about [shock] structure; . . .''~\cite[p.~361]{Schmidt69}.  In Section~\ref{sect:Q-metric} (below), we use $Q$ to  quantify   the degree of asymmetry exhibited by each of the four  $g$ vs.\ $\zeta$ profiles studied below.

Lastly, we define, following Morro~\cite{Morro06} and Straughan~\cite{S08}, the amplitude of the  jump in a function $\mathfrak{F}=\mathfrak{F}(\zeta)$ across the plane $\zeta =\zeta_{\rm d}$ as
\be\label{eq:Jump_def}
\lshad \mathfrak{F}\rshad
\Big{|}_{\zeta_{\rm d}} =\lim_{\zeta \to\zeta_{\rm d}^-} \mathfrak{F}(\zeta) - \lim_{\zeta \to\zeta_{\rm d}^+} \mathfrak{F}(\zeta),
\en
where it is assumed that both limits exist and that they are different. Here, we call attention to the fact that $\lshad \mathfrak{F}\rshad$ is positive (resp.~negative) when the jump in $\mathfrak{F}$ is from higher (resp.~lower) to lower (resp.~higher) values.

\section{Constant shear viscosity case}

Recall that under Case~(i), $\mu=\mu_{0}$; consequently, Eq.~\eqref{eq:TWmotion-final} becomes 
\be \label{eq:TWmotion-i}
(4\nu_{0}/3)(v - f) f^{\prime}+ vu_{\rm p}(1 - f/u_{\rm p})f =0,
\en
 which on setting $\mathcal{F}(\zeta) = -1+(2/u_{\rm p})f(\zeta)$ is reduced to
\be \label{eq:TWmotion_F-i}
(k +  \mathcal{F}) \mathcal{F}^{\prime} = \frac{3v}{4\nu_{0}}(1 - \mathcal{F}^{2}).
\en
This ODE, like the former, is a particularly simple  special case of Abel's equation~\cite{Davis62}; as such, it is easily integrated and yields the exact, but generally implicit\footnote{Except for certain values of $\textrm{Ma}$ that yield  explicit expressions; see Appendix~\ref{App:B}.}, TWS:
\begin{multline}\label{eq:TWS-i}
\left(\frac{3v}{4\nu_{0}}\right)\zeta 
= k \tanh^{-1}(\mathcal{F})-\frac{1}{2}\ln(1-\mathcal{F}^{2}) \qquad ( |\mathcal{F}| <1).
\end{multline}
Here, the constant of integration is zero, by way of the fact that $f(0)=u_{\rm p}/2$ implies $\mathcal{F}(0) =0$; we have set
\be
k = 1 - 2v/u_{\rm p} = -\sqrt{1+4 \textrm{Ma}^{-2}}; 
\en
and we note for later reference (in Appendix~\ref{App:B}) that Eq.~\eqref{eq:TWS-i} can also be expressed as
\begin{multline}\label{eq:TWS-Ln-Ln-i4}
\left(\frac{3v}{2\nu_{0}}\right)\zeta =(k-1)\ln(1+\mathcal{F}) 
- (k+1)\ln(1-\mathcal{F}) \qquad ( |\mathcal{F}| <1).
\end{multline}

Because $k <-1$,  the integral curves described by Eq.~\eqref{eq:TWS-i} can only take the form of fully dispersed shocks, also referred  to by some as \emph{kinks}; see, e.g., Ref.~\cite[\S\,5.2.2]{MATCOM16}.  For this velocity traveling wave profile, therefore, we can show that the shock thickness is given by
\be
\ell_{\rm i}=\frac{8(\nu_{0}/b_{0})\textrm{Ma}}{3\!\left(\textrm{Ma}+\sqrt{4+\textrm{Ma}^{2}}\,\right)\left( - 2 +\sqrt{4+\textrm{Ma}^{2}}\,\right)},
\en
to which corresponds the stationary point
\begin{eqnarray*}
\zeta_{\rm i}^{\bullet} =-\left(\frac{4\nu_{0}}{3v}\right)\Bigg{[}\sqrt{1+4\textrm{Ma}^{-2}}
 \times \tanh^{-1}\left(\frac{-2+\sqrt{4+\textrm{Ma}^{2}}}{\textrm{Ma}}\,\right)\\
+\frac{1}{2}\ln\left(\frac{-8+4\sqrt{4+\textrm{Ma}^{2}}}{\textrm{Ma}^{2}}\,\right)\Bigg{]};
\end{eqnarray*}
here, we observe that 
\be\label{eq:CalF-prime-max}
\mathcal{F}(\zeta_{\rm i}^{\bullet})= \mathcal{F}_{\rm i}^{\bullet} :=|k|-\sqrt{k^{2} - 1}=|k|-2/\textrm{Ma},
\en
which when related back to $f$ yields
\be\label{eq:f-prime-max}
f(\zeta_{\rm i}^{\bullet}) = f_{\rm i}^{\bullet} := \tfrac{1}{2}u_{\rm p}\left(1+|k|-2/\textrm{Ma}\right).
\en

In order to determine $l_{\rm i}$, we must first determine the (only) positive root of $\Pi(Y)=0$, where 
\be
\Pi(Y)=Y^{2} + 2|k|Y - 3.
\en
This quadratic arises when we attempt to solve $g^{\prime \prime}(\zeta)  =0$ after expressing it in terms of $\mathcal{F}$ and simplifying, i.e., when seeking the solution of
\be
[3(1 - \mathcal{F}^{2}) +2\mathcal{F}(k +  \mathcal{F})]\mathcal{F}^{\prime}=0,
\en
which we do subject to the constraint  $|\mathcal{F}|\in (0,1)$.   Denoting the aforementioned root by $Y=\mathcal{F}_{\rm i}^{*}$, it is not difficult to establish that
\be\label{eq:CalFi}
\mathcal{F}(\zeta_{\rm i}^{*})=\mathcal{F}_{\rm i}^{*} := - |k|  +\sqrt{k^{2} + 3},
\en
which when related back to $f$ yields
\be
f(\zeta_{\rm i}^{*}) = f_{\rm i}^{*} := \tfrac{1}{2}u_{\rm p}\left(1- |k|  +\sqrt{k^{2} + 3}\,\right)\!,
\en
where  the  stationary point of the corresponding    $g^{\prime}$ vs.\ $\zeta$ profile is  given by
\begin{equation}\label{eq:zeta_star-i}
\zeta_{\rm i}^{*} = \left[\frac{8\nu_{0}}{3u_{\rm p}(|k|+1)}\right]\! \left\{ k \tanh^{-1}(\mathcal{F}_{\rm i}^{*})-\tfrac{1}{2}\ln\left[1-(\mathcal{F}_{\rm i}^{*})^{2}\right]\right\}\!.
\end{equation}
With Eq.~\eqref{eq:CalFi} in hand and observing  that, under Case~(i), Eq.~\eqref{eq:g-prime} becomes
\begin{equation} \label{eq:TWg_prime-i}
 g^{\prime}(\zeta)  =  -\frac{3\rho_{0} v^2(u_{\rm p}f-f^2)}{4\nu_{0}(v - f)^{3}} 
 =\left(\frac{3v^{2}\rho_{0}}{2u_{\rm p}\nu_{0}}\right)\!\frac{(1 - \mathcal{F}^{2})}{(k +  \mathcal{F})^{3}}, 
\end{equation}
the density profile is seen to admit the shock thickness
\be
l_{\rm i}=  \frac{16(\nu_{0}/b_{0})}{3\textrm{Ma}(|k|-1)(|k|+1)^2} \left\{\frac{|k +  \mathcal{F}_{\rm i}^{*}|^{3}}{1-(\mathcal{F}_{\rm i}^{*})^{2}}\right\}.
\en

\noindent
\textbf{Remark 2:} Small-$|\zeta|$ and large-$|\zeta|$ approximations to $\mathcal{F}$ can easily  be determined from the corresponding expressions for ``$\mathcal{U}$'' given in Ref.~\cite[Remark~9]{MATCOM16}, wherein ``$\xi$'' plays  the role of $\zeta$. Mention should also be made  of the explicit, but approximate,  result~\cite[Remark~10]{MATCOM16}
\be\label{eq:Case-i_approx}
f(\zeta) \approx \tfrac{1}{2}u_{\rm p}\left[1-\tanh\left(2\zeta/\hat{\ell}_{\rm i} \right) \right] \qquad (\textrm{Ma} \ll 1),
\en
for which the shock thickness   is  $\hat{\ell}_{\rm i}= 16\nu_{0}/(3b_{0}\textrm{Ma})$, and we note that  $\textrm{Ma} \ll 1$ implies $|k|\gg 1$.

\section{The case $\mu \propto \rho$}

Under Case~(ii), $\mu=\nu_{0}\times\,$Eq.~\eqref{eq:TWg}; as such, Eq.~\eqref{eq:TWmotion-final} is reduced to the following Bernoulli-type equation:
\be \label{eq:TWmotion-ii}
(4\nu_{0}/3) f^{\prime}+ u_{\rm p}(1 - f/u_{\rm p})f =0,
\en
which is easily  integrated and yields the exact TWS
\be\label{eq_f_ii}
f(\zeta)=\frac{u_{\rm p}}{1+\exp\left(4\zeta/\ell_{\rm ii}\right)},
\en
where the shock thickness for this case of $f$ is given by
\be\label{eq: ell_ii}
\ell_{\rm ii}=\frac{16(\nu_{0}/b_{0})}{3\,\textrm{Ma}} \qquad \qquad \qquad (\zeta_{\rm ii}^{\bullet}=0).
\en
With the aid of Eq.~\eqref{eq_f_ii}, the density profile for this case is found to be
\begin{equation}\label{eq_g_ii}
g(\zeta)=  \rho_{0}\left(1+\sqrt{1+4 \textrm{Ma}^{-2}}\,\right)%\\
\times \!\left[1+\sqrt{1+4 \textrm{Ma}^{-2}}-\frac{2}{1+\exp \left(4\zeta/\ell_{\rm ii}\right)}\right]^{-1}\!;
\end{equation}
 and this profile admits the shock thickness 
\be\label{w}
l_{\rm ii}=\ell_{\rm ii},
\en
with corresponding stationary point
\be
\zeta_{\rm ii}^{*} =\left(\frac{\ell_{\rm ii}}{4}\right)\ln\left( \frac{-1+\sqrt{1+4 \textrm{Ma}^{-2}}\,}{1+\sqrt{1+4 \textrm{Ma}^{-2}}}\,\right)\!.
\en

From the latter it is readily established that $\zeta_{\rm ii}^{*}  < 0$, which follows from the fact that $\textrm{Ma} >0$, and, moreover, that 
\be
\zeta_{\rm ii}^{*} \approx  -\left(\frac{\ell_{\rm ii}\textrm{Ma}}{4}\right)\left(1-\frac{\textrm{Ma}^{2}}{24} \right) \qquad (\textrm{Ma} \ll 1),
\en
which  follows from the Taylor expansion of the $\ln$-term about $\textrm{Ma} = 0$.

\section{von Neumann--Richtmyer artificial viscosity}

Under this  formulation [i.e., Case~(iii)], $\mu$ is given by Eq.~\eqref{eq:vNR_mu};   Eq.~\eqref{eq:TWmotion-final}, therefore, becomes 
\be \label{eq:TWmotion-iii}
 \left(\frac{4\mathfrak{b}_{\rm iii}^{2}\lambda_{0}^{2}}{3}\right)\!(f^{\prime})^{2} - u_{\rm p}(1 - f/u_{\rm p})f =0,
\en
or the equivalent
\be\label{eq:Assoc_ODE_vNR_pm}
 \frac{\rd f}{\rd \zeta} = \mp \left(\frac{\sqrt{3}}{2\mathfrak{b}_{\rm iii}\lambda_{0}}\right)\! \sqrt{f \left(u_{\rm p}-  f \right)}.
\en

A phase plane analysis of Eq.~\eqref{eq:Assoc_ODE_vNR_pm}  reveals that its  equilibrium solutions  $\{0, u_{\rm p}\}$ are rendered stable and unstable, respectively, and thus consistent with the compressive version of the  piston problem, \emph{only} when the ``$-$'' sign case is selected.  

On \emph{rejecting} the ``$+$'' sign case, it is a straightforward matter  to show  (see, e.g., Refs.~\cite{MV12,vNR50}) that the velocity TWS under the vNR case is given by the following piecewise defined integral curve  of Eq.~\eqref{eq:Assoc_ODE_vNR_pm}:
\be\label{eq:TWS_vNR_f}
f(\zeta)=\begin{cases}
u_{\rm p}, & \zeta \leq - \tfrac{\pi}{4}\ell_{\rm iii},\\
\displaystyle{\tfrac{u_{\rm p}}{2}\left[1-\sin (2\zeta/\ell_{\rm iii})\right],} & -\tfrac{\pi}{4}\ell_{\rm iii} < \zeta < \tfrac{\pi}{4}\ell_{\rm iii},\\
0, & \zeta \geq \tfrac{\pi}{4}\ell_{\rm iii},
\end{cases}
\en
the shock thickness of which is [recall Eq.~\eqref{eq:ell_gen_f}] 
\be\label{eq:ell_iii}
\ell_{\rm iii}= \frac{4\lambda_{0}\mathfrak{b}_{\rm iii}}{\sqrt{3}} \qquad \qquad  (\zeta_{\rm iii}^{\bullet}=0).
\en
We also find, on substituting Eq.~\eqref{eq:TWS_vNR_f} into Eq.~\eqref{eq:TWfg} and making use of Eq.~\eqref{eq:v}, that the density TWS for this case  is given by
\begin{equation}\label{eq:TWS_vNR_g}
g(\zeta) = \rho_{0}%\\
\times\!\begin{cases}
\frac{1+\sqrt{1+4 \textrm{Ma}^{-2}}}{-1+\sqrt{1+4 \textrm{Ma}^{-2}}}, & \zeta \leq - \tfrac{\pi}{4}\ell_{\rm iii},\\
\frac{1+\sqrt{1+4\textrm{Ma}^{-2}}}{\sin (2\zeta/\ell_{\rm iii}) + \sqrt{1+4\textrm{Ma}^{-2}}}, & -\tfrac{\pi}{4}\ell_{\rm iii} < \zeta < \tfrac{\pi}{4}\ell_{\rm iii},\\
1, & \zeta \geq \tfrac{\pi}{4}\ell_{\rm iii},
\end{cases}
\end{equation}
 to which corresponds the (density) shock thickness   [recall Eq.~\eqref{eq:l_gen_g}]
 \begin{equation}
l_{\rm iii}=\ell_{\rm iii}\!\left(\frac{\textrm{Ma}}{4\sqrt{2}}\right)%\\
\times \!\left[\frac{20+9\textrm{Ma}^{2}-3\sqrt{(4+\textrm{Ma}^{2})(4+9\textrm{Ma}^{2})}}{\sqrt{-4-3\textrm{Ma}^{2}+\sqrt{(4+\textrm{Ma}^{2})(4+9\textrm{Ma}^{2})}}}\right]\!.
\end{equation}
The stationary point corresponding to $l_{\rm iii}$ is given by
\be
\zeta_{\rm iii}^{*} = \tfrac{1}{2}\ell_{\rm iii}\,\sin^{-1} \!\left(\!\frac{ \sqrt{1+4\textrm{Ma}^{-2}}-\sqrt{9+4\textrm{Ma}^{-2}}}{2}\,\right)\!,
\en
which we note can also be expressed as
\begin{equation}
\zeta_{\rm iii}^{*} =-\tfrac{1}{2}\ell_{\rm iii} %\\
\times \!\cos^{-1} \!\left(\!\sqrt{\frac{-3-4\textrm{Ma}^{-2}+\sqrt{16\textrm{Ma}^{-4}+40\textrm{Ma}^{-2}+9}}{2}}\,\right)\!,
\end{equation}
 where we observe that  $-\tfrac{\pi}{4}\ell_{\rm iii}< \zeta_{\rm iii}^{*} < 0$.\\

 \noindent
\textbf{Remark 3:}  The $f$ vs.\ $\zeta$ profile admits a pair of \emph{weak discontinuities}\footnote{Here, we use the terminology of Bland~\cite[p.~182]{Bland88}.}, both of second order; i.e., the $f^{\prime\prime}$ vs.\ $\zeta$ profile exhibits
two jumps, the amplitudes and locations of which are:
\be
\lshad f^{\prime\prime}\,\rshad
\Big{|}_{-\tfrac{\pi}{4}\ell_{\rm iii}} = \quad  \lshad f^{\prime\prime}\,\rshad
\Big{|}_{+ \tfrac{\pi}{4}\ell_{\rm iii}} = \quad 2u_{\rm p}/\ell_{\rm iii}^{2}.
\en
Likewise,  the present $g$ vs.\ $\zeta$ profile  also  exhibits (two) weak discontinuities of order two; in the case of these jumps we have
\begin{equation}
\lshad g^{\prime\prime}\,\rshad
\Big{|}_{\mp \tfrac{\pi}{4}\ell_{\rm iii}} =    \frac{4\rho_{0}\left(1+\sqrt{1+4 \textrm{Ma}^{-2}}\right)}{\ell_{\rm iii}^{2}\left(1-\sqrt{1+4 \textrm{Ma}^{-2}}\right)^{2}},  % \\
\frac{4\rho_{0}}{\ell_{\rm iii}^{2}\left(1+\sqrt{1+4 \textrm{Ma}^{-2}}\right)},
\end{equation}
respectively.

\section{Evans--Harlow--Longley artificial viscosity}\label{sect:Case_iv_fg}

Under this formulation [i.e., Case~(iv)],  for which $\mu$ is given by Eq.~\eqref{eq:EH_mu}, Eq.~\eqref{eq:TWmotion-final} reduces to
\be \label{eq:TWmotion-iv}
f\left[\left(\frac{2\mathfrak{b}_{\rm iv}\lambda_{0}}{3}\right)\! f^{\prime} + u_{\rm p}(1 - f/u_{\rm p})\right] =0,
\en
which like Eq.~\eqref{eq:TWmotion-i} is a special case of Abel's equation.  The integration of this ODE is not difficult; omitting the details, we find that
\be\label{eq:TWS_f_iv}
f(\zeta) = \begin{cases}
u_{\rm p}\left[1-(1/2)\exp(\zeta/\ell_{\rm iv})\right]\!, & \zeta < \ell_{\rm iv}\ln(2),\\
0, & \zeta\geq \ell_{\rm iv}\ln(2).\\
\end{cases}
\en
Here, the corresponding shock thickness is given by
\be
\ell_{\rm iv} =(2/3)\lambda_{0}\mathfrak{b}_{\rm iv},
\en
which we computed by evaluating the limit\footnote{Made necessary by the fact that, due to the discontinuity exhibited by $f^{\prime}$ under this case (see Remark~4), Eq.~\eqref{eq:ell_gen_f} is not applicable.}
\be
\lim_{\zeta \to (\zeta_{\rm iv}^{\bullet})^{-}}\frac{u_{\rm p}}{|f^{\prime}(\zeta)|}  \quad \qquad \qquad (\zeta_{\rm iv}^{\bullet} =\ell_{\rm iv}\ln 2).
\en

Similarly, we have for the  density profile
\be\label{eq:TWS_g_iv}
g(\zeta)=\rho_{0}\begin{cases}
\frac{1+\sqrt{1+4 \textrm{Ma}^{-2}}}{\exp(\zeta/\ell_{\rm iv})-1+\sqrt{1+4 \textrm{Ma}^{-2}}}, & \zeta < \ell_{\rm iv}\ln(2),\\
1, & \zeta\geq \ell_{\rm iv}\ln(2),\\
\end{cases}
\en
which admits the  shock thickness
\be
l_{\rm iv}= \frac{8\ell_{\rm iv}}{1+\sqrt{1+4 \textrm{Ma}^{-2}}} = \frac{16\mathfrak{b}_{\rm iv}\lambda_{0}}{3\left(1+\sqrt{1+4 \textrm{Ma}^{-2}}\right)},
\en
with
\be
\zeta_{\rm iv}^{*}=\ell_{\rm iv}\ln\left( -1+\sqrt{1+4 \textrm{Ma}^{-2}}\,\right)\!.
\en

\noindent
\textbf{Remark 4:}  As one differentiation of  Eq.~\eqref{eq:TWS_f_iv} reveals, the $f$ vs.\ $\zeta$ profile under this case exhibits an acoustic \emph{acceleration wave}\footnote{See, e.g., Ref.~\cite[\S\,8.1.3]{S08}, as well as those cited therein;  such a wave is also referred to by some as a ``first order weak discontinuity''~\cite[p.~182]{Bland88} and  a ``discontinuity wave''~\cite{BS14}.}; in other words, the $f^{\prime}$ vs.\ $\zeta$ profile suffers a jump discontinuity, the amplitude and location of which are:
\be
\lshad f^{\prime}\,\rshad
\Big{|}_{\ell_{\rm iv}\ln(2)} =  - u_{\rm p}/\ell_{\rm iv}.
\en
From Eq.~\eqref{eq:TWS_g_iv} it is clear that the $g$ vs.\ $\zeta$ profile also exhibits an acceleration wave, whose amplitude and location are 
\be
\lshad g^{\prime}\,\rshad
\Big{|}_{\ell_{\rm iv}\ln(2)} =    \frac{-2(\rho_{0}/\ell_{\rm iv})\textrm{Ma}}{\textrm{Ma}+\sqrt{4+\textrm{Ma}^{2}}}.
\en

\section{Numerical results}\label{sect:Num}

\subsection{Parameter values: Ar}\label{sect:param-values}

In the case of Ar at $\vartheta_{0} = 300\,\textrm{K}$ and $p_{0}=50\,\text{mTorr}$, the  gas/conditions on/under which Alsmeyer's~\cite{Alsmeyer76} shock experiments were performed, we have the following:
\begin{multline}
 \rho_{0} = 0.00010676\,\text{kg/m$^{3}$},\quad   \mu_{0}=2.2656\times 10^{-5}\,\text{Pa\,$\cdot$\,s}, \\   
 c_{0}= 322.59\,\text{m/s}, \quad 
 \end{multline}
from which we find that 
\begin{multline}\label{eq:ParaNum2}
b_{0}\approx 249.88\,\text{m/s}, \quad  \nu_{0}\approx 0.212\,\text{m$^{2}$/s}, \quad \lambda_{0}\approx 1.084\,\text{mm}.  
\end{multline}
Here, we have also made use of the result $\gamma =5/3$, which according to the kinetic theory of gases holds  for  all monatomic gases; see Refs.~\cite{T72,ChapCow70,TM80}.     The values of  $\rho_{0}$,  $\mu_{0}$, and $c_{0}$ were obtained from  the \emph{NIST Chemistry WebBook, SRD 69} 
(see: \url{https://webbook.nist.gov/chemistry/form-ser/}); those of $b_{0}$,  $\nu_{0}$, and $\lambda_{0}$, in contrast,  were computed using Eq.~\eqref{eq:b_0}, the defining relation $\nu_{0}=\mu_{0}/\rho_{0}$,  and Eq.~\eqref{eq:MFP_nu0_b0},  respectively.

And so as to achieve $\ell_{\rm iii}=\ell_{\rm iv}=3\lambda_{0}$, which follows from Ref.~\cite[p.~233]{Roache72},  based on our assumption $\Delta x =\lambda_{0}$ and the  shock thickness  definition we have adopted,   $\mathfrak{b}_{\rm iii}=\tfrac{1}{4}\sqrt{27}\,$ and $\mathfrak{b}_{\rm iv}= 9/2$ shall henceforth be taken. 

\subsection{Shock thickness results}\label{sect:delta_shock-thick}

\begin{figure*}
\begin{center}
  \includegraphics[width=0.75\textwidth]{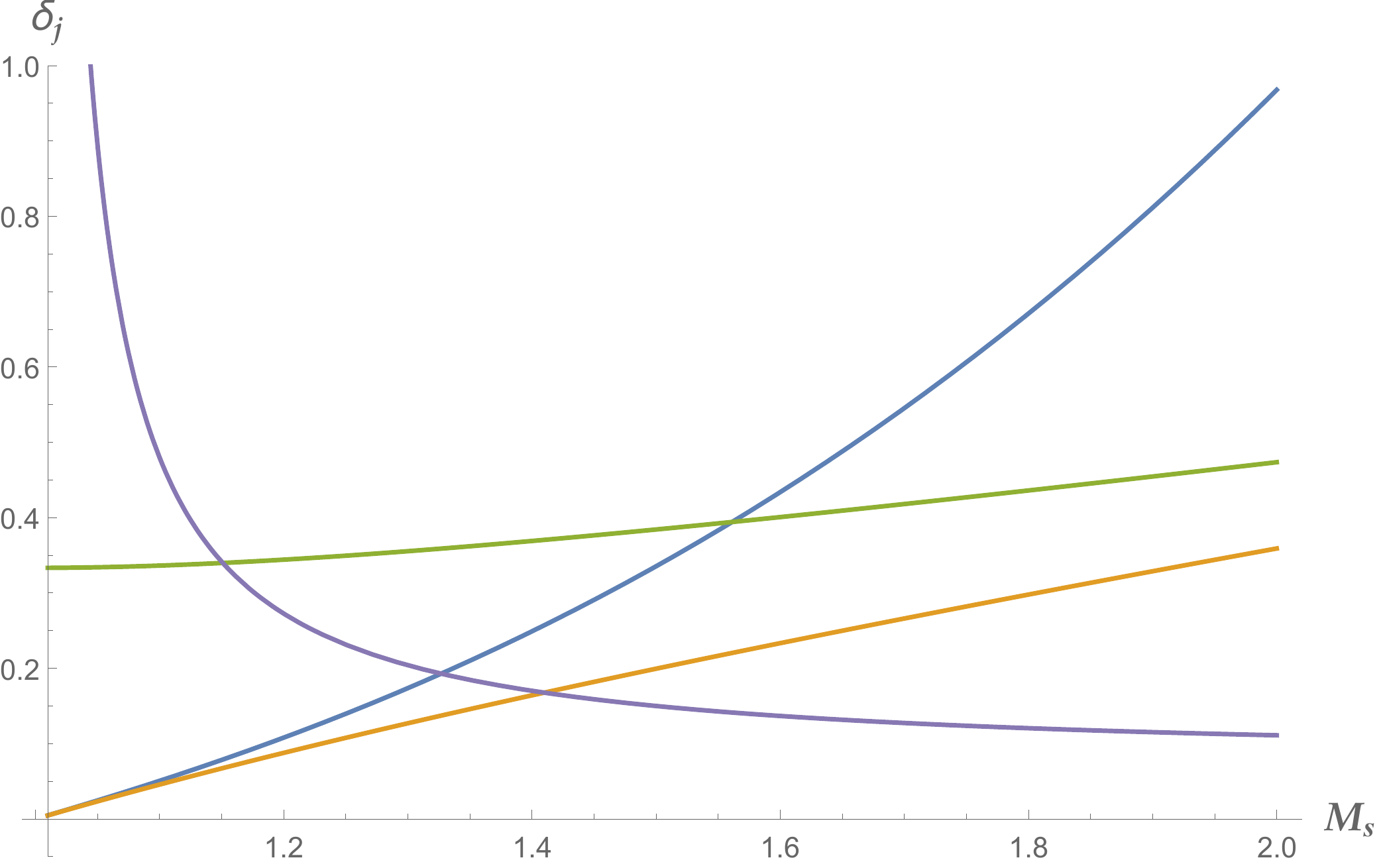}
\caption{$\delta_{j}$ vs.\ $M_{\rm s}$, based on Ref.~\cite[Fig.~2]{Alsmeyer76}, for the gas Ar. Blue:~Case~(i). Orange:~Case~(ii). Green:~Case~(iii). Lavender:~Case~(iv).}
\label{fig:delta}       
\end{center}
\end{figure*}

Fig.~\eqref{fig:delta} has been generated in accordance with Ref.~\cite[Fig.~2]{Alsmeyer76}, which  displays data for \emph{non-isothermal} shock propagation  in Ar.  In doing so, we have introduced the  dimensionless reciprocal shock thickness parameter, viz.:
\be
\delta_{j}=\lambda_{0}\left(l_{j}\Big{|}_{\textrm{Ma}=(M_{\rm s}^{2}-1)/M_{\rm s}}\right)^{-1} \quad (\text{$j =\,$i, ii, iii, iv}),
\en
and the \emph{shock} Mach number $M_{\rm s} = v/b_{0}$, where we observe that  $M_{\rm s} > 1$.  

Fig.~\eqref{fig:delta} reveals that, of the four curves we have plotted,  only the one corresponding to Case~(ii) exhibits qualitative agreement with Ref.~\cite[Fig.~2]{Alsmeyer76}.  The quantitative disagreement, however, between this curve and the data in Ref.~\cite[Fig.~2]{Alsmeyer76}, i.e., the fact that the former predicts a smaller shock thickness than the latter, is not unexpected. This is because under the isothermal assumption, $K$ is absent from the NSF system [recall Sys.~\eqref{sys:NS_Iso}]; i.e., the  dissipation associated  with the  gas's ability to conduct heat, which  tends to increase the shock's thickness, does \emph{not} occur in isothermal propagation. 

\subsection{Determination of $Q$-metric values}\label{sect:Q-metric}

In Table~\ref{table:Q}, values of $Q$ based on  the  parameter values given above for  Ar are presented.  While those for Cases~(ii)--(iv), as well as the $M_{\rm s}=\sqrt{2}$ case of Case~(i) (see Appendix~\ref{App:B}), were computed using expressions obtained from  evaluating the integrals in Eq.~\eqref{eq:Q} analytically, the entries  corresponding to the remaining $M_{\rm s}$ values under Case~(i) were computed from the numerical solution of Eq.~\eqref{eq:TWmotion-i}, the process of which involved  use of  both the \texttt{NDSolve[\,\,]} and \texttt{NIntegrate[\,\,]} commands provided in \textsc{Mathematica}.

\begin{table}[htbp]
\caption{Values of $Q$ in Ar corresponding to Cases~(i)--(iv) for selected Mach number  values}\label{table:Q}
\begin{minipage}{10cm}
\begin{threeparttable}[t]
  %\begin{center}
    \begin{tabular}{c|c|c|c|c} 
      Case & $\underset{(\textrm{Ma}=0.01)}{M_{\rm s}\approx 1.005}$ & $\underset{(\textrm{Ma} =1/\sqrt{2})}{M_{\rm s}=\sqrt{2}}$  & $\underset{(\textrm{Ma}=2/\sqrt{3})}{M_{\rm s}=\sqrt{3}}$ & $\underset{(\textrm{Ma}=3/2)}{M_{\rm s}=2}$\\
      \hline
      (i) & 1.010 & 2 & 3.000 & 4.000  \\
      \hline
     (ii) & 1.015 & 2.710 & 4.819 & 7.213 \\
     \hline
      (iii)\tnote{\dag} & 1.014 & 2.585 & 4.464 & 6.535 \\
      \hline
      (iv)\tnote{\ddag} & 2.625 & 6.885 & 12.239 & 18.378 \\
      \hline
    \end{tabular}
    \begin{tablenotes}
    \item[\dag]For this case, $\mp\infty$ in Eq.~\eqref{eq:Q} have been changed to $\mp\tfrac{1}{4}\pi\ell_{\rm iii}$, respectively.
    \item[\ddag]For this case, $+\infty$ in Eq.~\eqref{eq:Q} has been changed to $\ell_{\rm iv}\ln(2)$.
    \end{tablenotes}
    \end{threeparttable}
    \end{minipage}
  %\end{center}
\end{table}
From Table~\ref{table:Q} it is clear that, in all cases shown, $Q$ is a strictly increasing function of the shock Mach number; i.e., profile asymmetry increases with $M_{\rm s}$.  This behavior, we observe, is in agreement with the data presented in Refs.~\cite{Schmidt69,Alsmeyer76} for  non-isothermal shock propagation in Ar.  It is also clear from Table~\ref{table:Q} that Cases~(i)--(iii) all yield values of $Q$ that are quite close to unity (i.e., quite close to being perfectly symmetric),  and to each other, for $\textrm{Ma} = 0.01$.  In this regard, we point out that $\textrm{Ma} \ll 1$ defines the ``realm'' of finite-amplitude (or weakly-nonlinear) acoustics theory, within which velocity traveling wave profiles  can be expected to have the form of Eqs.~\eqref{eq:Case-i_approx} and~\eqref{eq_f_ii}; see, e.g., Refs.~\cite{MATCOM16,MRJ17}, as well as those cited therein.

Finally, the last three entries in the first row of Table~\ref{table:Q} highlight the following noteworthy feature associated with Case~(i):
\begin{equation}\label{eq:Q_Case-i}
\int_{-\infty}^{0}[1-R(\zeta)]\,\rd \zeta = M_{\rm s}^{2}\int_{0}^{+\infty} R(\zeta)\,\rd \zeta, 
\end{equation}
for $M_{\rm s}=\sqrt{2}, \sqrt{3}, 2$.  Interestingly, as shown in Appendix~\ref{App:B}, 
these   values are three of the five shock Mach number values for which  Case~(i) yields explicit TWSs.

\subsection{Determination of $\textrm{Ma}_{\rm sup}$ values}\label{sect:Max-Mach}

Implicit in Eq.~\eqref{eq:Theta_ext-final} is the thermodynamic requirement $\Theta_{\rm e} >0$.   A simplified approach---which  usually yields only approximate\footnote{Except in the case of certain viscosity laws, for which exact results follow; see Sections~\ref{Sect:Case-ii_e} and~\ref{Sect:Case-iii_e} below.\label{ft:exact}} results, however---of ensuring that this inequality is everywhere satisfied begins with recasting Eq.~\eqref{eq:Theta_ext-final} as the inequality
\begin{equation}\label{eq:Theta_ext-max}
0< c_{\rm v} \varkappa_{0}\vartheta_{0}-b_{0}^{2}\max|f^{\prime}| % \\
-  (4/3)\max\left[\frac{\mu(v - f)\left|f^{\prime}\right|^{2}}{\rho_{0} v}\right]\!, 
\end{equation}
which is easily broken down  into the  individual cases
\begin{equation}\label{eq:Theta_ext-max-cases}
0<  c_{\rm v} \varkappa_{0}\vartheta_{0}  -   \begin{cases}\displaystyle{\frac{b_{0}^{2}u_{\rm p}}{\ell_{\rm i}} +\frac{4\nu_{0}}{3}\!\left(\frac{u_{\rm p}^{2}}{\ell_{\rm i}^{2}}\right)}, &\text{(i)},  %\\
\\
\displaystyle{\frac{b_{0}^{2}u_{\rm p}}{\ell_{\rm ii}} + \frac{4\nu_{0}}{3}\!\left(\frac{u_{\rm p}^{2}}{\ell_{\rm ii}^{2}}\right)}, &\text{(ii)},
\end{cases}
\end{equation}
\begin{small}
\begin{equation}\label{eq:Theta_ext-max-cases_AV}
0<  c_{\rm v} \varkappa_{0}\vartheta_{0}   -  \begin{cases}
   \begin{cases}
   \displaystyle{\frac{b_{0}^{2}u_{\rm p}}{\ell_{\rm iii}} +\frac{9\lambda_{0}^{2}}{4}\!\left(\frac{u_{\rm p}^{3}}{\ell_{\rm iii}^{3}}\right)}, & |\zeta | \in [0, \tfrac{\pi}{4}\ell_{\rm iii}),\\
   0, & |\zeta | \geq \tfrac{\pi}{4}\ell_{\rm iii},
   \end{cases}
   &\text{(iii)},\\
   \\
   \begin{cases}
\displaystyle{\frac{b_{0}^{2}u_{\rm p}}{\ell_{\rm iv}} + 3\lambda_{0}\!\left(\frac{u_{\rm p}^{3}}{\ell_{\rm iv}^{2}}\right)},   & \zeta < \ell_{\rm iv}\ln(2),\\
0,  &\zeta \geq \ell_{\rm iv}\ln(2),
\end{cases}
&\text{(iv)},
\end{cases}
\end{equation}
\end{small}
where, in deriving these inequalities, we have made use of Eq.~\eqref{eq:ell_gen_f}.   It is noteworthy  that, because $f$ is a \emph{semi-compact}\footnote{In the sense  of Destrade et al.~\cite{DGS07}.} function under our two artificial  viscosity laws,  both inequalities in 
Relation~\eqref{eq:Theta_ext-max-cases_AV} become simply $0 < \vartheta_{0}$ on those intervals of the $\zeta$-axis over which $f$ (and therefore $f^{\prime}$) is identically zero.

In the remainder of this subsection,  we seek the value of $\textrm{Ma}_{\rm sup}$  for each case appearing in Relations~\eqref{eq:Theta_ext-max-cases} and~\eqref{eq:Theta_ext-max-cases_AV}, assuming one exists, where $\textrm{Ma} \geq  \textrm{Ma}_{\rm sup}$ means that isothermal propagation is \emph{not} possible because the speed of the piston (i.e., $u_{\rm p}$) is too great relative to $b_{0}$.  The numerical values    given below were computed based on $\gamma =5/3$ and the simplifying assumption  $M_{\rm r}=1$, where we define   the \emph{radiative} Mach number as $M_{\rm r} := b_{0}^{-1}\sqrt{\varkappa_{0}\nu_{0}}$.

\subsubsection{Case~(i)}\label{Sect:Case-i_e}

In dimensionless form, this case of Relation~
\eqref{eq:Theta_ext-max-cases} can be recast as
$0 < \Pi_{\rm i}(\textrm{Ma})$, where
\begin{small}
\begin{multline}
\Pi_{\rm i}(\textrm{Ma})=\Omega_{\rm i}(M_{\rm r})- \left(\textrm{Ma}+\sqrt{4+\textrm{Ma}^{2}}\,\right)\!\left(- 2 +\sqrt{4+\textrm{Ma}^{2}}\,\right)  \\
-\tfrac{1}{2}\left[\left(\textrm{Ma}+\sqrt{4+\textrm{Ma}^{2}}\,\right)\!\left(- 2 +\sqrt{4+\textrm{Ma}^{2}}\,\right)\right]^{2}, 
\end{multline}
\end{small}
and where we have set $\Omega_{\rm i}(M_{\rm r}) =(8/3)M_{\rm r}^{2}/(\gamma -1)$.

On solving  $\Pi_{\rm i}(\textrm{Ma}_{\rm sup}^{0})= 0$, which is readily accomplished due to its  bi-quadratic form, we find that $\textrm{Ma}_{\rm sup}\approx 
\textrm{Ma}_{\rm sup}^{0}$, where
\be
\textrm{Ma}_{\rm sup}^{0} = \begin{cases}
\frac{2Y_{\rm i}-\sqrt{2Y_{\rm i}}|Y_{\rm i}-4|}{2(Y_{\rm i}-2)}, &   \Omega_{\rm i}(M_{\rm r}) \neq  4,\\
3/2,   &  \Omega_{\rm i}(M_{\rm r}) = 4;
\end{cases}
\en
here,  $Y_{\rm i} =-1+\sqrt{1+2\Omega_{\rm i}(M_{\rm r})}$, where we observe that  $\Omega_{\rm i}(M_{\rm r})=4$ 
yields $Y_{\rm i} =2$,  and we have introduced the ``dummy'' variable $\textrm{Ma}_{\rm sup}^{0}$.

Unfortunately, $\textrm{Ma}_{\rm sup}^{0}$ is not, generally speaking, a very accurate approximation to 
$\textrm{Ma}_{\rm sup}$ under this case.   This is due  to the fact that, in general, 
\be
\max\big{|}\Theta_{\rm e}(\zeta)\big{|} > \big{|}\Theta_{\rm e}(\zeta_{\rm i}^{\bullet})\big{|},
\en
i.e., $\zeta_{\rm i}^{\rm e}\neq \zeta_{\rm i}^{\bullet}$,  where $\Theta_{\rm e}^{\prime}(\zeta_{\rm i}^{\rm e})=0$.   One can, in 
principle, determine $\textrm{Ma}_{\rm sup}$ for this case exactly by setting $\Theta_{\rm e}^{\prime}(\zeta)=0$, after first using 
Eq.~\eqref{eq:TWmotion-i} to eliminate $f^{\prime}$ from  Eq.~\eqref{eq:Theta_ext-final} and then applying $\rd/\rd \zeta$ to 
the result. The need to  employ Ferrari's  technique to solve the resulting quartic, however, renders this approach  far too 
involved, algebraically speaking,  to be performed here.  Instead, we  turn once again to {\sc Mathematica's} \texttt{NDSolve[]} 
command to numerically evaluate  Eq.~\eqref{eq:Theta_ext-final} in order to determine bounds on the value of 
$\textrm{Ma}_{\rm 
sup}$. 

Omitting the  details, it is not  difficult  to show that under Case~(i):
\begin{equation}
 \textrm{Ma}_{\rm sup}^{0} = 3/2 < 1.64108 < \textrm{Ma}_{\rm sup}  %\\
 < 1.64109  \qquad (M_{\rm r}=1, \gamma = 5/3),
\end{equation}
where it should be noted that setting $\textrm{Ma}_{\rm sup} = 1.64109$ yields $\min[\Theta_{\rm e}(\zeta)] < 0$.

\subsubsection{Case~(ii)}\label{Sect:Case-ii_e}

When expressed in dimensionless form, after using Eq.~\eqref{eq: ell_ii} to eliminate $\ell_{\rm ii}$, 
this case of Eq.~\eqref{eq:Theta_ext-max-cases} becomes
$0 < \Pi_{\rm ii}(\textrm{Ma})$, where
\be\label{eq:Pi_ii}
\Pi_{\rm ii}(\textrm{Ma}) = \Omega_{\rm ii}(M_{\rm r}) -\textrm{Ma}^{2}-\tfrac{1}{4}\textrm{Ma}^{4},
\en
and where $\Omega_{\rm ii}(M_{\rm r}) =2\Omega_{\rm i}(M_{\rm r}) = (16/3)M_{\rm r}^{2}/(\gamma -1)$.  

Because $\Theta_{\rm e}^{\prime}(\zeta_{\rm ii}^{\bullet})=0$ for all $\textrm{Ma}>0$, where recall that $\zeta_{\rm ii}^{\bullet}=0$, solving $\Pi_{\rm ii}(\textrm{Ma}_{\rm sup})= 0$ will give us the \emph{exact} value of  $\textrm{Ma}_{\rm sup}$ for this case; recall Footnote~\ref{ft:exact}. Since $\Pi_{\rm ii}(\textrm{Ma}_{\rm sup})$ is a  bi-quadratic, it is easily established  that  the  only positive root  $\Pi_{\rm ii}(\textrm{Ma}_{\rm sup})= 0$ is 
\be
\textrm{Ma}_{\rm sup} =\sqrt{-2+2\sqrt{1+\Omega_{\rm ii}(M_{\rm r})}}.
\en

Thus,  under Case~(ii):
\be
\textrm{Ma}_{\rm sup} = 2 \qquad (M_{\rm r}=1, \gamma = 5/3).
\en

\subsubsection{Case~(iii)}\label{Sect:Case-iii_e}

In dimensionless form, this case of Eq.~\eqref{eq:Theta_ext-max-cases} can be recast as
$0 < \Pi_{\rm iii}(\textrm{Ma})$; here,
\be\label{eq:Pi_iii}
\Pi_{\rm iii}(\textrm{Ma})=\Omega_{\rm iii}(M_{\rm r})-\tfrac{1}{3}\textrm{Ma}-\tfrac{1}{12}\textrm{Ma}^{3},
\en
where the $\mathfrak{b}_{\rm iii}=\tfrac{1}{4}\sqrt{27}$ special case of Eq.~\eqref{eq:ell_iii} has been used to eliminate $\ell_{\rm ii}$. Also, we have set
 \be
 \Omega_{\rm iii}(M_{\rm r}) = \frac{16 M_{\rm r}^{2}}{(\gamma -1)\sqrt{50\pi}},
\en
from which we have eliminated $\lambda_{0}$ with the aid of  Eq.~\eqref{eq:MFP_nu0_b0}.

Because $\Theta_{\rm e}^{\prime}(\zeta_{\rm iii}^{\bullet})=0$ for all $\textrm{Ma}>0$, where recall that $\zeta_{\rm iii}^{\bullet}=0$, solving $\Pi_{\rm iii}(\textrm{Ma}_{\rm sup})= 0$ will give us the \emph{exact} value of  $\textrm{Ma}_{\rm sup}$ for this case; recall, again, Footnote~\ref{ft:exact}.  Since $\Pi_{\rm iii}(\textrm{Ma}_{\rm sup})$ is a  \emph{depressed} cubic, applying Cardano's formula is a relatively simple matter; omitting the details, we find that
\begin{multline}
\textrm{Ma}_{\rm sup} = \sqrt[3]{6\Omega_{\rm iii}(M_{\rm r})+2\sqrt{9\Omega_{\rm iii}^{2}(M_{\rm r})+\tfrac{16}{27}}} \\
+\sqrt[3]{6\Omega_{\rm iii}(M_{\rm r})-2\sqrt{9\Omega_{\rm iii}^{2}(M_{\rm r})+\tfrac{16}{27}}},
\end{multline}
which  is  the only positive root of $\Pi_{\rm iii}(\textrm{Ma}_{\rm sup})= 0$.

Thus,  under Case~(iii):
\be
\textrm{Ma}_{\rm sup} \approx 2.379 \qquad (M_{\rm r}=1, \gamma = 5/3).
\en

\noindent
\textbf{Remark 5:}   The  $\Theta_{\rm e}^{\prime}(\zeta)$ vs.\ $\zeta$ profile exhibits 
two jumps under Case~(iii), the amplitudes and locations of which are:
\be
\lshad \Theta_{\rm e}^{\prime}\,\rshad
\Big{|}_{- \tfrac{\pi}{4}\ell_{\rm iii}} = \lshad \Theta_{\rm e}^{\prime}\,\rshad
\Big{|}_{+ \tfrac{\pi}{4}\ell_{\rm iii}}= \frac{2b_{0}^{2}u_{\rm p}}{9c_{\rm v}\varkappa_{0}\lambda_{0}^{2}}. 
\en

\subsubsection{Case~(iv)}\label{Sect:Case-iv_e}

Because the Case~(iv) version of  Eq.~\eqref{eq:Theta_ext-final} exhibits explicit dependence on not only $f^{\prime}$ but also $f$, the corresponding case of Eq.~\eqref{eq:Theta_ext-max-cases_AV} will only yield an approximation to $\textrm{Ma}_{\rm sup}$---one that can be shown  to be  quite poor.  

Hence, let us  specialize  Eq.~\eqref{eq:Theta_ext-final} to Case~(iv) (see Section~\ref{sect:Case_iv_fg}), with $\mathfrak{b}_{\rm iv}= 9/2$,  and then replace $\Theta_{\rm e}(\zeta)$ with its  dimensionless counterpart $T_{\rm e}(\zeta)$; doing so yields, after simplifying,  the exact expression
 \begin{equation}\label{eq:Te_iv}
T_{\rm e}(\zeta) = \Omega_{\rm iv}(M_{\rm r}) % \\
- \begin{cases}
 \tfrac{1}{6} \textrm{Ma} \exp(\zeta/\ell_{\rm iv})\\
  \qquad + \tfrac{1}{12} \textrm{Ma}^{3} \exp(2\zeta/\ell_{\rm iv}) \\
   \qquad \qquad - \tfrac{1}{24} \textrm{Ma}^{3}  \exp(3\zeta/\ell_{\rm iv}) & \zeta < \zeta_{\rm iv}^{\bullet}, \\
0,  & \zeta \geq \zeta_{\rm iv}^{\bullet},
\end{cases}
\end{equation}
where $T_{\rm e}(\zeta) \propto \Theta_{\rm e}(\zeta)$ and $\Omega_{\rm iv}(M_{\rm r}) =\Omega_{\rm iii}(M_{\rm r})$.  
 
Although doing so is not simply a matter of   setting $T_{\rm e}^{\prime}(\zeta)=0$, it can be shown that $\zeta_{\rm iv}^{\rm e}$, the critical point of $T_{\rm e}(\zeta)$ under this case, is exactly given by 
\be\label{eq:zeta_iv_e}
\zeta_{\rm iv}^{\rm e}=
\begin{cases} 
\ell_{\rm iv}\ln[Y_{\rm iv}(\textrm{Ma})], & \textrm{Ma} > 1,\\
 \zeta_{\rm iv}^{\bullet} = \ell_{\rm iv}\ln(2), &\textrm{Ma} \leq  1,
\end{cases}
\en
where we have set
 \be
 Y_{\rm iv}(\textrm{Ma})= \frac{2\textrm{Ma}+2\sqrt{3+\textrm{Ma}^{2}}}{3\textrm{Ma}} \qquad (\textrm{Ma} > 1).
 \en
Here, we observe that   $T_{\rm e}^{\prime}(\zeta_{\rm iv}^{\rm e})=0$ \emph{only} for $\textrm{Ma} >  1$, where the corresponding stationary point of the $T_{\rm e}(\zeta)$ vs.\ $\zeta$ profile  is also its absolute minimum; for $\textrm{Ma} \leq  1$, in contrast, $\inf[T_{\rm e}(\zeta)] =\lim_{\zeta \to (\zeta_{\rm iv}^{\bullet})^{-}}T_{\rm e}(\zeta)$.

Consequently, we are led to consider the inequality $0 < \Pi_{\rm iv}(\textrm{Ma})$, where 
\begin{equation}\label{eq:exp_iv}
\Pi_{\rm iv}(\textrm{Ma})=\Omega_{\rm iv}(M_{\rm r})  %\\
- \begin{cases}
\tfrac{1}{6} \textrm{Ma} Y_{\rm iv}  + \tfrac{1}{12} \textrm{Ma}^{3} Y_{\rm iv}^{2}- \tfrac{1}{24} \textrm{Ma}^{3} Y_{\rm iv}^{3}, & \textrm{Ma} >1,\\
\tfrac{1}{3} \textrm{Ma}, & \textrm{Ma} \leq 1.
\end{cases}
\end{equation}
Eq.~\eqref{eq:exp_iv} follows  from Eq.~\eqref{eq:Te_iv}  on setting  $\Pi_{\rm iv}(\textrm{Ma}) = T_{\rm e}(\zeta_{\rm iv}^{\rm e})$, and as such contains no approximations. 

It is now possible to determine $\textrm{Ma}_{\rm sup}$ exactly by solving  $\Pi_{\rm iv}(\textrm{Ma}_{\rm sup})= 0$, 
which can be simplified to read
\begin{equation}\label{eq:Cubic_iv}
0= \begin{cases}
(243\Omega_{\rm iv}^{2}(M_{\rm r})-4) - 54 \Omega_{\rm iv}(M_{\rm r})\textrm{Ma}_{\rm sup} \\
\qquad - \textrm{Ma}_{\rm sup}^{2} - 12\Omega_{\rm iv}(M_{\rm r})\textrm{Ma}_{\rm sup}^{3}, &M_{\rm r} > M_{\rm r}^{\star},\\
\Omega_{\rm iv}(M_{\rm r})-\tfrac{1}{3} \textrm{Ma}_{\rm sup}, & M_{\rm r} \leq M_{\rm r}^{\star},
\end{cases}
\end{equation}
where we note the critical value
\be
M_{\rm r}^{\star} :=\frac{1}{2}\sqrt{\frac{5(\gamma-1)}{3}\sqrt{\frac{\pi}{8}}},
\en 
and we observe that $\Omega_{\rm iv}(M_{\rm r}^{\star})=1/3$. 

\begin{figure}[t]
  \includegraphics[width=0.90\textwidth]{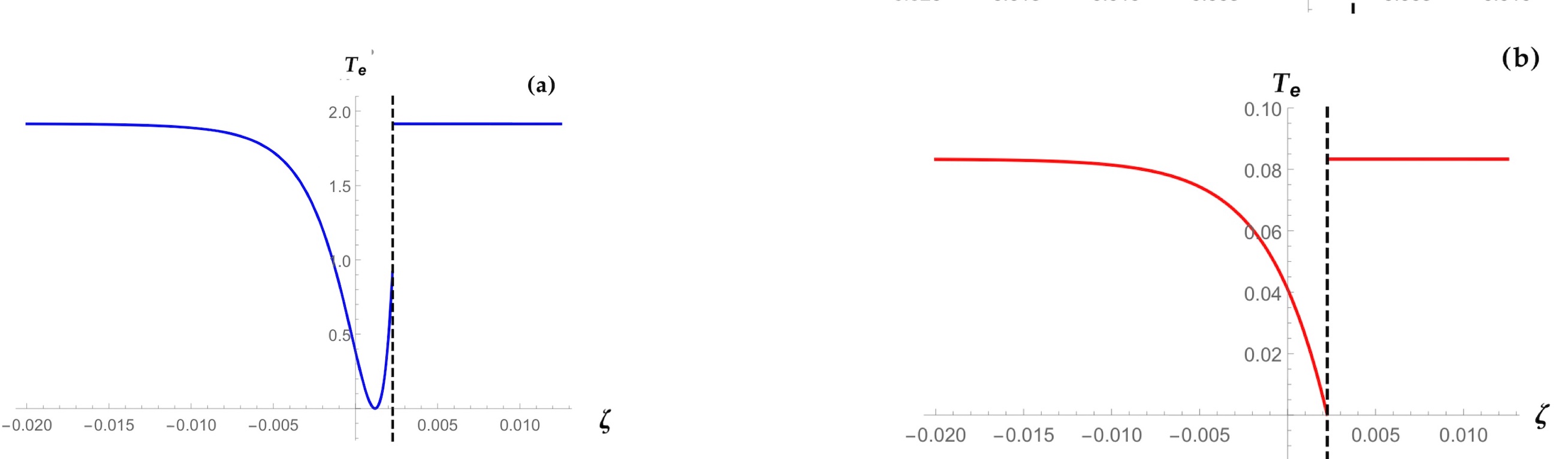}
\caption{$T_{\rm e}(\zeta)$ vs.\ $\zeta$ generated from Eq.~\eqref{eq:Te_iv}  using the parameter values for Ar given in Section~\ref{sect:param-values}, where we recall that  $\mathfrak{b}_{\rm iv}=9/2 \implies \ell_{\rm iv}=3\lambda_{0}$; here,  $M_{\rm r}^{\star}\approx 0.4172$. Blue curve:~$M_{\rm r} = 1$ (see Eq.~\eqref{eq:Case_iv_Mr1}). Red curve: $M_{\rm r}= \tfrac{1}{2}M_{\rm r}^{\star}$ ($\implies \textrm{Ma}_{\rm sup} = 1/4$). 
Black-broken line: $\zeta= \zeta_{\rm iv}^{\bullet}=3\lambda_{0}\ln(2)\approx 0.0023$.}
\label{fig:2}       
\end{figure}

While it is a trivial matter to establish that
\be
\textrm{Ma}_{\rm sup} = 3\Omega_{\rm iv}(M_{\rm r}) \qquad (M_{\rm r} \leq M_{\rm r}^{\star}),
\en
suffice it to say that the process of determining $\textrm{Ma}_{\rm sup}$ for the cubic case of Eq.~\eqref{eq:Cubic_iv}  is  
algebraically intensive.  As such, we set our analytical efforts aside and return to our numerical tools;  hence, following an 
approach similar to that used in Section~\ref{Sect:Case-i_e}, we find that  under Case~(iv): 
\begin{equation}\label{eq:Case_iv_Mr1}
  2.92641 < \textrm{Ma}_{\rm sup} 
 < 2.92642 \quad (M_{\rm r}=1, \gamma = 5/3),
\end{equation}
where we note that  setting $\textrm{Ma}_{\rm sup}= 2.92642$ yields\\ $\min[T_{\rm e}(\zeta)] < 0$.  And to help clarify the 
nature of the  absolute minimum it exhibits, we have  plotted $T_{\rm e}(\zeta)$ vs.\ $\zeta$ in Fig.~\ref{fig:2}(a,b), where the 
plots in the former and latter  correspond to $M_{\rm r} > M_{\rm r}^{\star}$ and $M_{\rm r} \leq M_{\rm r}^{\star}$, respectively. 
 \noindent
\textbf{Remark 6:}   The temperature of the surrounding environment  must exhibit a jump discontinuity across $\zeta=\zeta_{\rm iv}^{\bullet}$ under the present case; specifically (recall Eq.~\eqref{eq:Jump_def}) 
\be
\lshad T_{\rm e} \rshad \big{|}_{\zeta_{\rm iv}^{\bullet}} = (\Omega_{\rm iv}(M_{\rm r})  - \tfrac{1}{3}\textrm{Ma}) - \Omega_{\rm iv}(M_{\rm r}) = - \tfrac{1}{3}\textrm{Ma}, 
\en
for all $\textrm{Ma}>0$. The plots in Fig.~\ref{fig:2}(a,b) also serve to illustrate this phenomenon, with $\textrm{Ma}_{\rm sup}$  
in the role of $\textrm{Ma}$.    Evaluating numerically, we find that $\lshad T_{\rm e} \rshad \big{|}_{\zeta_{\rm iv}^{\bullet}} 
\approx  - 0.9755$ and $\lshad T_{\rm e} \rshad \big{|}_{\zeta_{\rm iv}^{\bullet}} \approx -0.0833$, respectively, 
in Fig.~\ref{fig:2}(a,b).

 \section{Closing remarks and observations}\label{sect:Close}
  
 \begin{enumerate}
 \item[$\bullet$] If $\textrm{Ma}=1$, then $v=b_{0}\varphi$, where $\varphi\approx 1.618$ is the \emph{proportio divina}~\cite{Livio02}; in other words, if $u_{\rm p}=b_{0}$, then 
 \be
 v\approx b_{0}\left(\frac{F_{N+1}}{F_{N}}\right) \qquad (N \gg 1), 
 \en
 where $F_{N}$ denotes the $N$th Fibonacci number; see Ref.~\cite[p.~101]{Livio02}. (Note that $F_{1}=F_{2}=1$.)
 
\item[$\bullet$] While $f, g \in \mathcal{C}^{\infty}(\mathbb{R})$ under both Cases~(i) and (ii), the same is not true under  Cases~(iii) and~(iv); specifically,  $f, g \in \mathcal{C}^{1}(\mathbb{R})$ and  $f, g \in \mathcal{C}^{0}(\mathbb{R})$, respectively, under the former and latter.

\item[$\bullet$] By averaging the (1D) Euler system under  finite-scale theory (FST), one obtains a system containing an ``effective viscosity''  that is given by Eq.~\eqref{eq:vNR_mu} with $\mathfrak{b}_{\rm iii}  =1/\sqrt{12}$  and $\lambda_{0}$ replaced by $L$, where $L(>0)$ is the length scale that appears in the averaging transform of  FST; see Refs.~\cite{MV12,MPR20} and those cited therein.

\item[$\bullet$] The $\Theta_{\rm e}(\zeta)$ vs.\ $\zeta$ profile under Case~(iii) admits a pair of \emph{temperature-rate waves} (see, e.g., Refs.~\cite{Morro06,S11,SZJ20}), i.e., jumps in $\Theta_{\rm e}^{\prime}(\zeta)$; see Remark~5. 

\item[$\bullet$]  A \emph{thermal shock} (see, e.g., Refs.~\cite{S11,SZJ20}), i.e., a jump in $T_{\rm e}(\zeta)$,  must occur under Case~(iv); see Remark~6.

\item[$\bullet$]  While they are quite distinct from those  presented for Ar in Refs.~\cite{Schmidt69,Alsmeyer76}, the TWS profiles generated under Case~(iv) \emph{are} very similar to those given in Ref.~\cite[Fig.~2]{JFM57}, which depict non-isothermal shocks in CO$_{2}$; see also Ref.~\cite{Taniguchi18}.
\end{enumerate}

\begin{center} \bf Acknowledgments\end{center}

All numerical computations and simulations presented above were performed using \textsc{Mathematica}, ver.~11.2.   S.C.\  wishes to acknowledge the financial  support of GNFM-INdAM, INFN, and \textsc{Sapienza}
University of Rome, Italy.   P.M.J.\  was supported by U.S.\ Office of Naval Research (ONR) funding.

\appendix

\section{Appendix: Comparison with Stokes~(1851)}\label{App:A}

%Here is an appendix. If there is more than one appendix the heading titles should be descriptive.
\renewcommand{\thesection}{A}
\setcounter{equation}{0}
\numberwithin{equation}{section}

In this appendix the EoM derived by Stokes in Ref.~\cite{Stokes51}, which describes non-isothermal propagation with radiation under the linear approximation, is re-derived.   We also give the isothermal version of Stokes' model and  the linearized, reduced, version of the present (i.e., isothermal) governing system.  Key aspects of all three models are  noted and briefly discussed.

In terms of the present notation, the governing system of equations considered by Stokes~\cite{Stokes51}, and later  Rayleigh~\cite[\S\,247]{Rayleigh96}\footnote{In Ref.~\cite[\S\,247(4)]{Rayleigh96}, the first two terms should be multiplied by $\kappa_{v}$, where $\kappa_{v}$ is used in  Ref.~\cite{Rayleigh96} to denote the specific heat at constant volume; see also Ref.~\cite[\S\,360(18)]{Lamb45}.}, can be expressed as
\begin{subequations}\label{sys:Linear_1}
\begin{align}
s_{t} &= - u_{x},\label{eq:Cont_s}\\
 \rho_{0}u_{t} &=-p_{x}\label{eq:Mom_u}\\
\theta_{t}-(\gamma-1)s_{t} & = - \varkappa_{0}\theta,\label{eq:Egy_eta}\\
p &= p_{0}(1+s+\theta). \label{eq:EoS_linear}
\end{align}
\end{subequations}
In Sys.~\eqref{sys:Linear_1},      $s=(\rho-\rho_0)/\rho_{0}$ is known as the condensation;  we have set  $\theta =(\vartheta-\vartheta_0)/\vartheta_{0}$; the term on the right-hand side of Eq.~\eqref{eq:Egy_eta} follows from taking, prior to cancellation of the product  $c_{\rm v} \rho_{0}\vartheta_{0}$,
\be\label{eq:Newton_lc}
r=- c_{\rm v} \varkappa_{0}\vartheta_{0}\theta,
\en
i.e.,  $r$ to be given by the general (i.e., non-isothermal) form of Newton's law of cooling;   and  $\mu$, $\mu_{\rm b}$, and $K$ have all been set equal to zero. (Unless stated herein, the definitions of all quantities, terms, etc., appearing in this appendix can be found in Section~\ref{sect:2}.)

Now  using Eqs.~\eqref{eq:Cont_s} and~\eqref{eq:EoS_linear} to eliminate $u$ and $p$, respectively, from Eq.~\eqref{eq:Mom_u} reduces Sys.~\eqref{sys:Linear_1}  to 
\begin{subequations}\label{sys:Linear_2}
\begin{align}
 s_{tt}  &= b_{0}^{2}s_{xx}+b_{0}^{2}\theta_{xx},\label{eq:Wave_s_temp}\\
(\varkappa_{0} + \partial_{t})\theta & = (\gamma-1)s_{t}, \label{eq:Egy_eta_relax}
 \end{align}
\end{subequations}
where  we have re-written Eq.~\eqref{eq:Egy_eta} in operator form.

At this point it is instructive to introduce the isothermal counterpart of Sys.~\eqref{sys:Linear_2}:
\begin{subequations}\label{sys:Linear_iso}
\begin{align}
 s_{tt}  - b_{0}^{2}s_{xx} &=0,\label{eq:Wave_linear-iso}\\
\vartheta_{\rm e} & =\vartheta_{0}[1- \varkappa_{0}^{-1}(\gamma-1)s_{t}], \label{eq:Egy_theta-e_iso}
 \end{align}
\end{subequations}
where, as in Section~\ref{sect:IsoPP},  $\vartheta_{\rm e}=\vartheta_{\rm e}(x,t)$ is controllable by the experimenter, and
\begin{subequations}\label{sys:Linear_iso-mu}
\begin{align}
 s_{tt}  - b_{0}^{2}s_{xx} &=\tfrac{4}{3}\nu_{0}s_{txx},\label{eq:Wave_linear-iso-mu}\\
\vartheta_{\rm e} & =\vartheta_{0}[1- \varkappa_{0}^{-1}(\gamma-1)s_{t}], \label{eq:Egy_theta-e_iso-mu}
 \end{align}
\end{subequations}
which  is easily obtained from the linearized version of Sys.~\eqref{sys:NS_Iso}.  Here, in parallel with the analyses carried out in Section~\ref{sect:Max-Mach}, we stress the fact that if $s_{t} >0$ (i.e., compression) is a possibility, then suitable additional restrictions must be placed on the  boundary and/or initial data for  Eqs.~\eqref{eq:Wave_linear-iso} and~\eqref{eq:Egy_theta-e_iso-mu}  to ensure that  Eqs.~\eqref{eq:Egy_theta-e_iso} and~\eqref{eq:Egy_theta-e_iso-mu} yield $\vartheta_{\rm e} > 0$.

In closing, we return to Sys.~\eqref{sys:Linear_2} and, on eliminating $\theta$ between its two PDEs, obtain  the modern form of the EoM first derived by Stokes in 1851, viz.,
\be\label{eq:SMGT}
s_{ttt}-c_{0}^{2}s_{txx} + \varkappa_{0}(s_{tt} - b_{0}^{2}s_{xx})  = 0;
\en
see Refs.~\cite[Eq.~(7)]{Stokes51} and~\cite[\S\,247(6)]{Rayleigh96}.  In recent years, many authors have begun referring to PDEs of this type as (Stokes--)Moore--Gibson--Thompson equations; see, e.g., Ref.~\cite[p.~3]{BV19}, wherein  an interesting stability condition for Eq.~\eqref{eq:SMGT} and its multi-D extensions is presented and discussed.  

With regard to modeling experiments involving both radiation and non-isothermal conditions, we observe that Sys.~\eqref{sys:Linear_2} is preferred to Eq.~\eqref{eq:SMGT} in terms of initial data requirements. That is,   while both formulations require knowledge of $s(x,0)$ and $s_{t}(x,0)$, it appears to be much easier, from the standpoint of performing the experiment,  for one to  accurately specify (and enforce) $\theta(x,0)$ than  $s_{tt}(x,0)$.

\appendix 
\renewcommand{\thesection}{B}
\section{Appendix: Explicit TWSs under Case~(i)}\label{App:B}
\setcounter{equation}{0}
\numberwithin{equation}{section}
We begin with the  observation that Eq.~\eqref{eq:TWS-Ln-Ln-i4} can also be written as
\begin{equation}\label{eq:TWS-Ln-Ln-i}
\frac{3v\zeta}{2\nu_{0}(|k|-1)} =  \ln(1-\mathcal{F}) %\\
 -\left(\frac{|k|+1}{|k|-1}\right)\ln(1+\mathcal{F}) \qquad ( |\mathcal{F}| <1),
\end{equation}
where we recall that $k <-1$.  Prompted by the special case TWS derived in Ref.~\cite[\S\,4]{MRJ17}, wherein   \emph{Becker's assumption} (see also Ref.~\cite{ML49}) was adopted, we find that setting $\textrm{Ma}=1/\sqrt{2}$ ($\implies k=-3$, $M_{\rm s}=\sqrt{2}$)  reduces Eq.~\eqref{eq:TWS-Ln-Ln-i} to
\be\label{eq:TWS_explicit_1st}
\left(\frac{3b_{0}\sqrt{2}}{4\nu_{0}}\right)\zeta =\ln\left[\frac{1-\mathcal{F}}{(1+\mathcal{F})^2} \right]\!,
\en
which is readily solved to yield the explicit expression 
\begin{equation}\label{eq:TWS_Bi}
f(\zeta)=\frac{1}{4}u_{\rm p}\exp\left[- (3+2\sqrt{2}\,)\zeta/\ell_{\rm i}\right]  %\\
\times \left\{-1+\sqrt{1+8\exp \left[ (3+2\sqrt{2}\,)\zeta/\ell_{\rm i}\right]}\,\right\}\!,
\end{equation}
where we recall our use of the relation $\mathcal{F}(\zeta) = -1+(2/u_{\rm p})f(\zeta)$. In the case of Eq.~\eqref{eq:TWS_Bi}, the shock thickness and corresponding stationary point are given by
\be
\ell_{\rm i} = \frac{\nu_{0}(3+2\sqrt{2})\sqrt{8}}{3b_{0}}, \,\,\,\, \zeta_{\rm i}^{\bullet}= \left(\frac{4\nu_{0}}{b_{0}\sqrt{18}}\right)\ln\!\left[\tfrac{1}{4}(1+\sqrt{2}\,)\right]\!, 
\en
respectively.

The corresponding density TWS  is easily found to be 
\begin{equation}\label{eq:TWS_rho_Bi}
g(\zeta)=\!\rho_{0}\Bigg{(}1-\tfrac{1}{8}\exp\left[- (3+2\sqrt{2}\,)\zeta/\ell_{\rm i}\right]  %\\
\times \left\{-1+\sqrt{1+8\exp \left[ (3+2\sqrt{2}\,)\zeta/\ell_{\rm i}\right]}\,\right\}\Bigg{)}^{-1}\!\!\!\!,
\end{equation}
and its shock thickness and stationary point are given by
\be
l_{\rm i} = 2(\nu_{0}/b_{0})\sqrt{6}, \qquad \zeta_{\rm i}^{*}= - l_{\rm i} \ln\left(2\right)\!.
\en

We close by pointing out  that \emph{four} other explicit  TWSs are possible under Case~(i).  Apart from noting that obtaining them requires one to solve the cubic and quartic equations
\be\label{eq:explicit_3_4}
(1-\mathcal{F})\exp\left[-\,\frac{3b_{0}\zeta\sqrt{m}}{2\nu_{0} (|k|-1)}\right] = (1+\mathcal{F})^{m},
\en
however, we leave their derivations to the reader; here,
\be\label{eq:explicit_m_k}
m=\begin{cases}
4/3 (\implies k=-7), & \textrm{Ma}=1/\sqrt{12},\\
3/2 (\implies k=-5), & \textrm{Ma}=1/\sqrt{6},\\
3 (\implies k=-2), & \textrm{Ma}=2/\sqrt{3},\\
4 (\implies k=-5/3), & \textrm{Ma}=3/2,
\end{cases}
\en
where $m(>1)$ is given by $m= (|k|+1)(|k|-1)^{-1}=M_{\rm s}^{2}$.

\end{document}